\documentclass[conference]{IEEEtran}

\usepackage{graphicx}
\usepackage[T1]{fontenc}
\usepackage[draft]{hyperref}

\usepackage{enumitem}
\usepackage[cmex10]{amsmath}
\usepackage{amssymb}
\usepackage{cite}
\usepackage{amsthm}
\usepackage{textcomp}
\usepackage{siunitx}
\usepackage{color}
\usepackage{subfigure}
\usepackage{cases}
\usepackage[font={small}]{caption}

\makeatletter
\def\thm@space@setup{\thm@preskip=2pt
\thm@postskip=2pt \itshape}
\makeatother
\newtheoremstyle{newstyle}      
{} 
{} 
{\mdseries} 
{} 
{\bfseries} 
{.} 
{ } 
{} 

\theoremstyle{newstyle}

\newtheorem{theorem}{Theorem}

\theoremstyle{definition}

\newtheorem{definition}{Definition}

\theoremstyle{remark}
\newtheorem{remark}{Remark}

\setlist[description]{style=multiline}

\let\emptyset\varnothing

\IEEEoverridecommandlockouts

\begin{document}
\sloppy

\setlength{\belowcaptionskip}{-6pt}
\setlength{\abovedisplayskip}{0.5mm}
\setlength{\belowdisplayskip}{0.5mm}
\setlength{\abovecaptionskip}{0.5mm}

\title{A Unified Coding Framework for Distributed Computing with Straggling Servers} 


\author{Songze~Li$^{*}$, Mohammad~Ali~Maddah-Ali $^{\dagger}$, and A.~Salman~Avestimehr$^{*}$\\
$^{*}$ Department of Electrical Engineering, University of Southern California, Los Angeles, CA, USA \\ 
$^{\dagger}$ Nokia Bell Labs, Holmdel, NJ, USA\\
}

\maketitle

\begin{abstract}
We propose a \emph{unified} coded framework for distributed computing with straggling servers, by introducing a tradeoff between ``latency of computation'' and ``load of communication'' for some linear computation tasks. We show that the coded scheme of~\cite{li2016fundamental,LMA_ISIT16,LMA_all} that repeats the intermediate computations to create coded multicasting opportunities to reduce communication load, and the coded scheme of~\cite{lee2015speeding,lee-ISIT16} that generates redundant intermediate computations to combat against straggling servers can be viewed as special instances of the proposed framework, by considering two extremes of this tradeoff:  minimizing either the load of communication or the latency of computation individually. Furthermore, the latency-load tradeoff achieved by the proposed coded framework allows to systematically operate at any point on that tradeoff to perform distributed computing tasks. We also prove an information-theoretic lower bound on the latency-load tradeoff, which is shown to be within a constant multiplicative gap from the achieved tradeoff at the two end points. 
\end{abstract}

\section{Introduction}\label{sec:intro}
Recently, there have been two novel ideas proposed to exploit coding in order to speed up distributed computing applications. Specifically, a repetitive structure of computation tasks across distributed computing servers was proposed in~\cite{li2016fundamental,LMA_ISIT16,LMA_all}, enabling coded multicast opportunities that significantly reduce the time to shuffle intermediate results.
On the other hand, applying Maximum Distance Separable (MDS) codes to some linear computation tasks (e.g., matrix multiplication) was proposed in~\cite{lee2015speeding,lee-ISIT16}, in order to alleviate the effects of straggling servers and shorten the computation phase of distributed computing.

In this paper, we propose a \emph{unified} coded framework for distributed computing with straggling servers, by introducing a tradeoff between ``latency of computation'' and ``load of communication'' for linear computation tasks. We show that the coding schemes of~\cite{li2016fundamental} and~\cite{lee2015speeding} can then be viewed as special instances of the proposed coding framework by considering two extremes of this tradeoff:  minimizing either the load of communication or the latency of computation individually. Furthermore, the proposed coding framework provides a natural tradeoff between computation latency and communication load in distributed computing, and allows to systematically operate at any point on that tradeoff.

More specifically, we focus on a distributed matrix multiplication problem in which for a matrix ${\bf A}$ and $N$ input vectors ${\bf x}_1,\ldots,{\bf x}_N$, we want to compute $N$ output vectors ${\bf y}_1={\bf A}{\bf x}_1,\ldots,{\bf y}_N={\bf A}{\bf x}_N$. The computation cannot be performed on a single server node since its local memory is too small to hold the entire matrix ${\bf A}$. Instead, we carry out this computation using $K$ distributed computing servers collaboratively. Each server has a local memory, with the size enough to store up to equivalent of $\mu$ fraction of the entries of the matrix {\bf A}, and it can only perform computations based on the contents stored in its local memory. Matrix multiplication is one of the building blocks to solve data analytics and machine learning problems (e.g., regression and classification). Many such applications of big data analytics require massive computation and storage power over large-scale datasets, which are nowadays provided collaboratively by clusters of computing servers, using efficient distributed computing frameworks such as Hadoop MapReduce~\cite{dean2004mapreduce} 
and Spark~\cite{zaharia2010spark}. Therefore, optimizing the performance of distributed matrix multiplication is of vital importance to improve the performance of the distributed computing applications.

A distributed implementation of matrix multiplication proceeds in three phases: Map, Shuffle and Reduce. In the Map phase, every server multiplies the input vectors with the locally stored matrix that partially represents the target matrix ${\bf A}$. When a subset of servers finish their local computations such that their Map results are sufficient to recover the output vectors, we halt the Map computation and start to Shuffle the Map results across the servers in which the final output vectors are calculated by specific Reduce functions.

Within the above three-phase implementation, the coding approach of \cite{li2016fundamental} targets at minimizing the shuffling load of intermediate Map results. 
It introduces a particular repetitive structure of Map computations across the servers, and utilizes this redundancy to enable a specific type of network coding in the Shuffle phase (named coded multicasting) to minimize the communication load. We term this coding approach as ``Minimum Bandwidth Code''. In~\cite{li2016scalable,LQMA_globecom16}, the Minimum Bandwidth Code was employed in a fully decentralized wireless distributed computing framework, achieving a scalable architecture with a constant load of communication. The other coding approach of~\cite{lee2015speeding}, however, aims at minimizing the latency of Map computations by encoding the Map tasks 
using MDS codes, so that the run-time of the Map phase is not affected by up to a certain number of straggling servers. This coding scheme, which we term as ``Minimum Latency Code'', results in a significant reduction of Map computation latency.

\begin{figure}[htbp]
   \centering
   \includegraphics[width=0.35\textwidth]{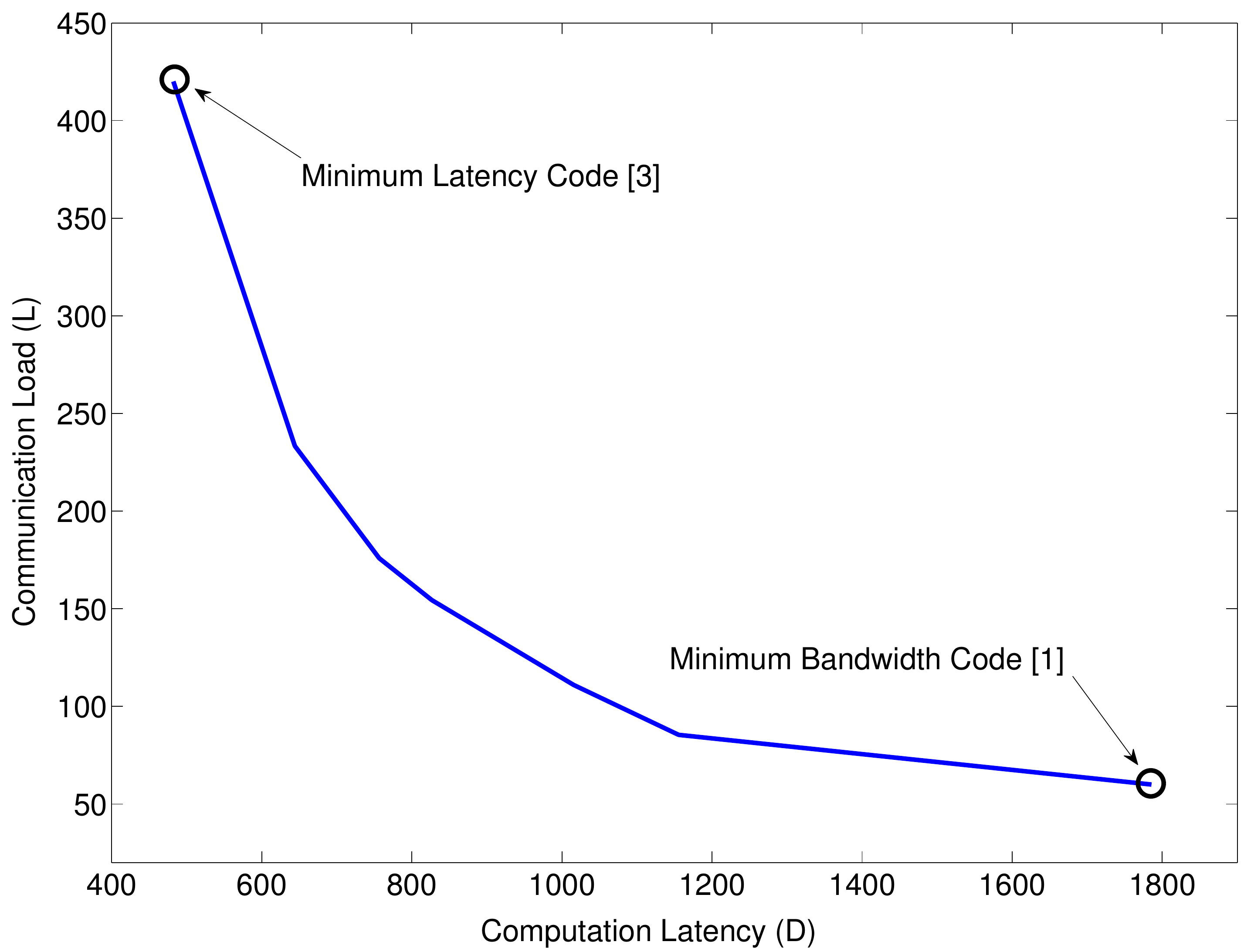}  
   \caption{The Latency-Load tradeoff, for a distributed matrix multiplication job of computing $N=840$ output vectors using $K=14$ servers each with a storage size $\mu=1/2$.}
   \label{fig:tradeoff}
\end{figure}

In this paper, we formalize a \emph{tradeoff} between the computation latency in the Map phase (denoted by $D$) and the communication (shuffling) load in the Shuffle phase (denoted by $L$) for distributed matrix multiplication (in short, the \emph{Latency-Load Tradeoff}), in which as illustrated in Fig.~\ref{fig:tradeoff}, the above two coded schemes correspond to the two extreme points that minimize $L$ and $D$ respectively. Furthermore, we propose a unified coded scheme that organically integrates both of the coding techniques, and allows to systematically operate at any point on the introduced tradeoff.

For a given computation latency, we also prove an information-theoretic lower bound on the minimum required communication load to accomplish the distributed matrix multiplication. This lower bound is proved by first concatenating multiple instances of the problem with different reduction assignments of the output vectors, and then applying the cut-set bound on subsets of servers. 
At the two end points of the tradeoff, the proposed scheme achieves the minimum communication load to within a constant factor. 

We finally note that there has been another tradeoff between the computation load in the Map phase and the communication load in the Shuffle phase for distributed computing, which is introduced and characterized in~\cite{li2016fundamental}. In this paper, we are fixing the amount of computation load (determined by the storage size) at each server, and focus on characterizing the tradeoff between the computation latency (determined by the number of servers that finish the Map computations) and the communication load. Hence, the considered tradeoff can be viewed as an extension of the tradeoff in~\cite{li2016fundamental} by introducing a third axis, namely the computation latency of the Map phase.

\section{Problem Formulation}\label{sec:def}
\subsection{System Model}
We consider a matrix multiplication problem in which given a matrix ${\bf A} \in \mathbb{F}_{2^T}^{m \times n}$ for some integers $T$, $m$ and $n$, and $N$ input vectors ${\bf x}_1,\ldots,{\bf x}_N \in \mathbb{F}_{2^T}^n$, we want to compute $N$ output vectors ${\bf y}_1 = {\bf A}{\bf x}_1,\ldots,{\bf y}_N = {\bf A}{\bf x}_N$.

We perform the computations using $K$ distributed servers. Each server has a local memory of size $\mu mnT$ bits (i.e., it can store equivalent of $\mu$ fraction of the entries of the matrix ${\bf A}$), for some $\frac{1}{K} \leq \mu \leq 1$.\footnote{Thus enough information to recover the entire matrix ${\bf A}$ can be stored collectively on the $K$ servers.}

We allow applying linear codes for storing the rows of ${\bf A}$ at each server. Specifically, Server $k$, $k \in \{1,\ldots,K\}$, designs an encoding matrix ${\bf E}_k \in \mathbb{F}_{2^T}^{\mu m \times m}$, and stores
\begin{equation}\label{eq:store}
{\bf U}_k = {\bf E}_k {\bf A}.
\end{equation}
The encoding matrices ${\bf E}_1,\ldots,{\bf E}_K$ are design parameters and is denoted as \emph{storage design}. The storage design is performed in prior to the computation.

\begin{remark}
For the Minimum Bandwidth Code in~\cite{li2016fundamental}, each server stores $\mu m$ rows of the matrix ${\bf A}$. Thus, the rows of the encoding matrix ${\bf E}_k$ was chosen as a size-$\mu m$ subset of the rows of the identity matrix ${\bf I}_m$, according to a specific repetition pattern. While for the Minimum Latency Code in~\cite{lee2015speeding}, ${\bf E}_k$ was generated randomly such that every server stores $\mu m$ random linear combinations of the rows of ${\bf A}$, achieving a $(\mu m K, m)$ MDS code. $\hfill \square$
\end{remark}

\vspace{-2.5mm}
\subsection{Distributed Computing Model}	
\vspace{-1.5mm}
	We assume that the input vectors ${\bf x}_1,\ldots,{\bf x}_N$ are known to all the servers. The overall computation proceeds in three phases: \emph{Map}, \emph{Shuffle}, and \emph{Reduce}.
	
	\noindent {\bf Map Phase:} The role of the Map phase is to compute some coded intermediate values according to the locally stored matrices in (\ref{eq:store}), which can be used later to re-construct the output vectors. More specifically, for all $j=1,\ldots,N$, Server $k$, $k =1,\ldots,K$, computes the intermediate vectors
	\begin{equation}\label{eq:map}
	{\bf z}_{j,k} = {\bf U}_k {\bf x}_j = {\bf E}_k {\bf A}{\bf x}_j = {\bf E}_k{\bf y}_j.
	\end{equation} 

We denote the latency for Server~$k$ to compute ${\bf z}_{1,k},\ldots,{\bf z}_{N,k}$ as $S_k$. We assume that $S_1,\ldots,S_K$ are i.i.d. random variables, and denote the $q$th order statistic, i.e., the $q$th smallest variable of $S_1,\ldots,S_K$ as $S_{(q)}$, for all $q \in \{1,\ldots,K\}$. We focus on a class of distributions of $S_k$ such that
	\begin{align}
	\mathbb{E}\{S_{(q)}\} = \mu N g(K,q),
	\end{align}
for some function $g(K,q)$.

	


The Map phase terminates when a subset of servers, denoted by ${\cal Q} \subseteq \{1,\ldots,K\}$, have finished their Map computations in (\ref{eq:map}). A necessary condition for selecting ${\cal Q}$ is that the output vectors ${\bf y}_1\ldots,{\bf y}_N$ can be re-constructed by jointly utilizing the intermediate vectors calculated by the servers in ${\cal Q}$, i.e., $\{{\bf z}_{j,k}: j=1,\ldots,N, k \in {\cal Q}\}$. However, one can allow redundant computations in ${\cal Q}$, since if designed properly, they can be used to reduce the load of communicating intermediate results, for servers in ${\cal Q}$ to recover the output vectors in the following stages of the computation.


\begin{remark}
The Minimum Bandwidth Code in~\cite{li2016fundamental} waits for all servers to finish their computations, i.e., ${\cal Q}=\{1,\ldots,K\}$. For the Minimum Latency Code in~\cite{lee2015speeding}, ${\cal Q}$ is the subset of the fastest $\lceil \frac{1}{\mu}\rceil$ servers in performing the Map computations. $\hfill \square$
\end{remark}

\begin{definition}[Computation Latency]
We define the \emph{computation latency}, denoted by $D$, as the average amount of time spent in the Map phase. $\hfill\Diamond$
\end{definition}

After the Map phase, the job of computing the output vectors ${\bf y}_1\ldots,{\bf y}_N$ is continued \emph{exclusively} over the servers in ${\cal Q}$. 
The final computations of the output vectors are distributed uniformly across the servers in ${\cal Q}$. We denote the set of indices of the output vectors assigned to Server $k$ as ${\cal W}_k$, and $\{{\cal W}_k: k\in {\cal Q}\}$ satisfy 1) ${\cal W}_k \cap {\cal W}_{k'} = \emptyset, \; \forall k \neq k'$, 2) $|{\cal W}_k| = N/|{\cal Q}|\in \mathbb{N}, \; \forall k \in {\cal Q}$.\footnote{We assume that $N \gg K$, and $|{\cal Q}|$ divides $N$ for all ${\cal Q} \subseteq \{1,\ldots,K\}$.}
	
\noindent {\bf Shuffle Phase:} The goal of the Shuffle phase is to exchange the intermediate values calculated in the Map phase, to help each server recover the output vectors it is responsible for. To do this, every server $k$ in ${\cal Q}$ generates a message $X_k$ from the locally computed intermediate vectors ${\bf z}_{1,k},\ldots,{\bf z}_{N,k}$ through an encoding function $\phi_k$, i.e., 
$X_k = \phi_k\left({\bf z}_{1,k},\ldots,{\bf z}_{N,k}\right)$,
such that upon receiving all messages $\{X_k: k \in {\cal Q}\}$, every server $k \in {\cal Q}$ can recover the output vectors in ${\cal W}_k$. We assume that the servers are connected by a shared bus link. After generating $X_k$, Server~$k$ multicasts $X_k$ to all the other servers in ${\cal Q}$. 
	
\begin{definition}[Communication Load]
	We define the \emph{communication load}, denoted by $L$, as the average total number of bits in all messages $\{X_k: k \in {\cal Q}\}$, normalized by $mT$ (i.e., the total number of bits in an output vector). $\hfill\Diamond$
\end{definition}

\noindent {\bf Reduce Phase:} The output vectors are re-constructed distributedly in the Reduce phase. 
Specifically, User $k$, $k \in {\cal Q}$, uses the locally computed vectors ${\bf z}_{1,k},\ldots,{\bf z}_{N,k}$ and the received multicast messages $\{X_k: k \in {\cal Q}\}$ to recover the output vectors with indices in ${\cal W}_k$ via a decoding function $\psi_k$, i.e.,
\begin{align}
\{{\bf y}_j: j \in {\cal W}_k\} = \psi_k({\bf z}_{1,k},\ldots,{\bf z}_{N,k},\{X_k: k \in {\cal Q}\}).
\end{align}
	
For such a distributed computing system, we say a latency-load pair $(D,L) \in \mathbb{R} ^2$ is \emph{achievable} if there exist a storage design $\{{\bf E}_k\}_{k=1}^K$, a Map phase computation with latency $D$, and a shuffling scheme with communication load $L$, such that all output vectors can be successfully reduced. 

\begin{definition}
We define the latency-load region, as the closure of the set of all achievable $(D,L)$ pairs.     $\hfill \Diamond$
\end{definition}

\subsection{Illustrating Example}\label{sec:illustrate-example}
In order to clarify the formulation, we use the following simple example to illustrate the latency-load pairs achieved by the two coded approaches discussed in Section~\ref{sec:intro}.

We consider a matrix ${\bf A}$ consisting of $m=12$ rows ${\bf a}_1,\ldots,{\bf a}_{12}$. We have $N=4$ input vectors ${\bf x}_1,\ldots,{\bf x}_4$, and the computation is performed on $K=4$ servers each has a storage size $\mu =\frac{1}{2}$. We assume that the Map latency $S_k$, $k=1,\ldots,4$, has a shifted-exponential distribution function 
	\begin{equation}\label{eq:dis}
	F_{S_k}(t) = 1-e^{-(\frac{t}{\mu N}-1)}, \; \forall t \geq \mu N,
	\end{equation}
and by e.g.,~\cite{arnold1992first}, the average latency for the fastest $q$, $1\leq q \leq 4$, servers to finish the Map computations is
\begin{equation}
	D(q)=\mathbb{E}\{S_{(q)}\} = \mu N\Big(1 + \sum_{j=K-q+1}^{K} \tfrac{1}{j}\Big).
\end{equation}

\begin{figure}[htbp]
   \centering
\subfigure[Minimum Bandwidth Code. Every row of ${\bf A}$ is multiplied with the input vectors twice. For $k =1,2,3,4$, Server $k$ reduces the output vector ${\bf y}_k$. In the Shuffle phase, each server multicasts $3$ bit-wise XORs, denoted by $\oplus$, of the calculated intermediate values, each of which is simultaneously useful for two other servers. \vspace{-2mm}]{\includegraphics[width=0.48\textwidth]{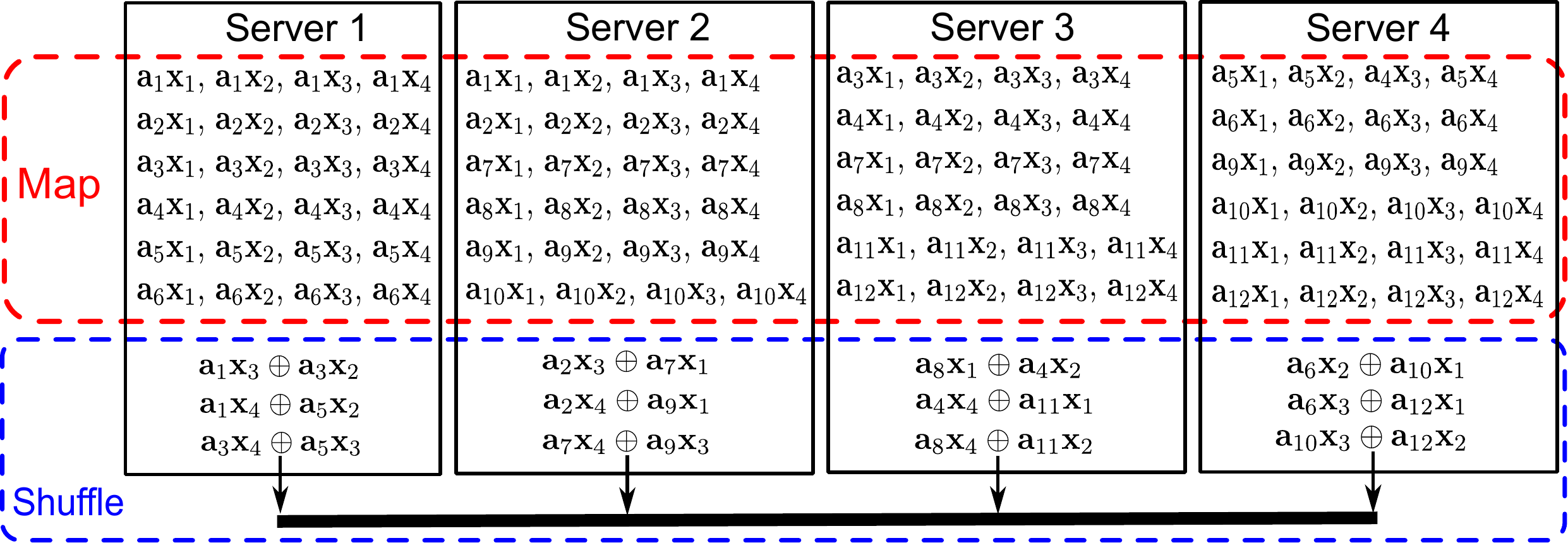}
       \label{fig:shuffle}}
       \vspace{-1.5mm}
 \subfigure[Minimum Latency Code. ${\bf A}$ is encoded into 24 coded rows ${\bf c}_1\ldots,{\bf c}_{24}$. Server 1 and 3 finish their Map computations first. They then exchange enough number (6 for each output vector) of intermediate values to reduce ${\bf y}_1, {\bf y}_2$ at Server~1 and ${\bf y}_3, {\bf y}_4$ at Server~3.]{\includegraphics[width=0.48\textwidth]{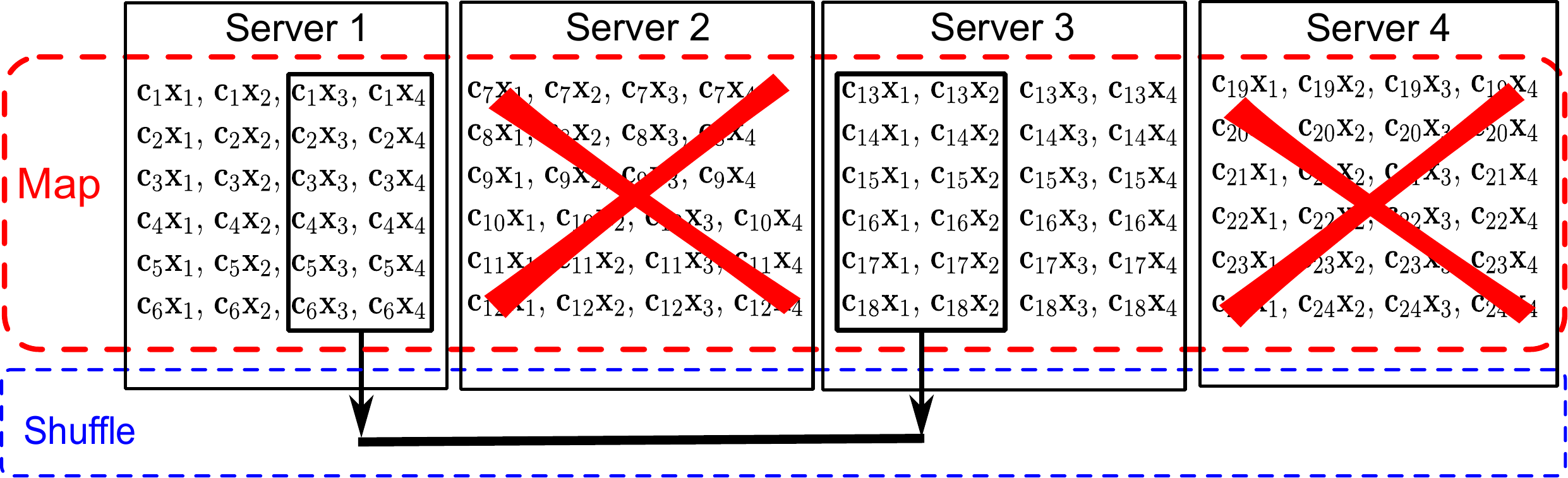}
       \label{fig:map}}
   \caption{Illustration of the Minimum Bandwidth Code in~\cite{li2016fundamental} and the Minimum Latency Code in~\cite{lee2015speeding}.}
   \label{fig:extreme}
    \vspace{-2.5mm}
\end{figure}

\noindent {\bf Minimum Bandwidth Code~\cite{li2016fundamental}.} The Minimum Bandwidth Code in~\cite{li2016fundamental}  repeatedly stores each row of ${\bf A}$ at $\mu K$ servers with a particular pattern, such that in the Shuffle phase, $\mu K$ required intermediate values can be delivered with a single coded multicast message, which results in a coding gain of $\mu K$. We illustrate such coding technique in Fig.~\ref{fig:shuffle}.

As shown in Fig.~\ref{fig:shuffle}, a Minimum Bandwidth Code repeats the multiplication of each row of ${\bf A}$ with all input vectors ${\bf x}_1,\ldots,{\bf x}_4$, $\mu K=2$ times across the $4$ servers, e.g., ${\bf a}_1$ is multiplied at Server~1 and~2. 
The Map phase continues until all servers have finished their Map computations, achieving a computation latency $D(4)=2\times(1+\sum_{j=1}^4 \frac{1}{j})=\frac{37}{6}$. For $k=1,2,3,4$, Server $k$ will be reducing output vector ${\bf y}_k$. In the Shuffle phase, as shown in Fig.~\ref{fig:shuffle}, due to the specific repetition of Map computations, every server multicasts $3$ bit-wise XORs, each of which is simultaneously useful for two other servers. For example, upon receiving ${\bf a}_1{\bf x}_3 \oplus {\bf a}_3{\bf x}_2$ from Server 1, Server 2 can recover $ {\bf a}_3{\bf x}_2$ by canceling ${\bf a}_1{\bf x}_3$ and Server 3 can recover $ {\bf a}_1{\bf x}_3$ by canceling ${\bf a}_3{\bf x}_2$. Similarly, every server decodes the needed values by canceling the interfering values using its local Map results. The Minimum Bandwidth Code achieves a communication load $L = 3 \times 4/12=1$.

The Minimum Bandwidth Code can be viewed as a specific type of network coding~\cite{ahlswede2000network}, or more precisely index coding~\cite{birk2006coding,bar2011index}, in which the key idea is to design ``side information'' at the servers (provided by the Map results), enabling multicasting opportunities in the Shuffle phase to minimize the communication load. 

\noindent {\bf Minimum Latency Code~\cite{lee2015speeding}.} The Minimum Latency Code in~\cite{lee2015speeding} uses MDS codes to generate some redundant Map computations, and assigns the coded computations across many servers. Such type of coding takes advantage of the abundance of servers so that one can terminate the Map phase as soon as enough coded computations are performed across the network, without needing to wait for the remaining straggling servers. We illustrate such coding technique in Fig.~\ref{fig:map}.


For this example, a Minimum Latency Code first has each server $k$, $k=1,\ldots,4$, independently and randomly generate $6$ random linear combinations of the rows of ${\bf A}$, denoted by ${\bf c}_{6(k-1)+1},\ldots,{\bf c}_{6(k-1)+6}$ (see Fig.~\ref{fig:map}).
We note that $\{{\bf c}_1,\ldots,{\bf c}_{24}\}$ is a $(24,12)$ MDS code of the rows of ${\bf A}$. Therefore, for any subset ${\cal D} \subseteq \{1,\ldots,24\}$ of size $|{\cal D}|=12$, using the intermediate values $\{{\bf c}_i{\bf x}_j: i \in {\cal D}\}$ can recover the output vector ${\bf y}_j$. The Map phase terminates once the fastest $2$ servers have finished their computations (e.g., Server~1 and~3), achieving a computation latency $D(2)\!=\! 2\! \times \!(1+\frac{1}{3}+\frac{1}{4})\!=\!\frac{19}{6}$. Then Server~1 continues to reduce ${\bf y}_1$ and ${\bf y}_2$, and Server~3 continues to reduce ${\bf y}_3$ and ${\bf y}_4$. As illustrated in Fig.~\ref{fig:map}, Server~1 and~3 respectively unicasts the intermediate values it has calculated and needed by the other server to complete the computation, 
achieving a communication load $L \!=\! 6\! \times \! 4/12\!=\!2$.



From the above descriptions, we note that the Minimum Bandwidth Code uses about twice of the time in the Map phase compared with the Minimum Latency Code, and achieves half of the communication load in the Shuffle phase. They represent the two end points of a general latency-load tradeoff characterized in the next section.

\section{Main Results}
The main results of the paper are, 1) a characterization of a set of achievable latency-load pairs by developing a unified coded framework, 2) an outer bound of the latency-load region, which are stated in the following two theorems. 

\vspace{-1mm}
\begin{theorem}
For a distributed matrix multiplication problem of computing $N$ output vectors using $K$ servers, each with a storage size $\mu \geq \frac{1}{K}$, the latency-load region contains the lower convex envelop of the points
\begin{align}
\{(D(q),L(q)): q =\lceil \tfrac{1}{\mu}\rceil,\ldots,K\},\label{eq:pair}
\end{align}
in which 
\begin{align}
D(q) &= \mathbb{E}\{S_{(q)}\} = \mu N g(K,q),\label{eq:latency}\\
L(q) &= N\sum_{j=s_q}^{\lfloor \mu q \rfloor} \tfrac{B_j}{j} + N\min\big\{1-\bar{\mu}-\sum_{j=s_q}^{\lfloor \mu q \rfloor} B_j, \tfrac{B_{s_q-1}}{s_q-1}\big\}, \label{eq:load}
\end{align}
where $S_{(q)}$ is the $q$th smallest latency of the $K$ i.i.d. latencies $S_1,\ldots,S_K$ with some distribution $F$ to compute the Map functions in (\ref{eq:map}), $g(K,q)$ is a function of $K$ and $q$ computed from $F$, $\bar{\mu} \triangleq \frac{\lfloor \mu q\rfloor}{q}$, $B_j \triangleq \frac{{q-1 \choose j}{K-q \choose \lfloor \mu q \rfloor-j}}{\frac{q}{K} {K \choose \lfloor \mu q \rfloor}}$, and $s_q \triangleq \inf \{s: \sum_{j=s}^{\lfloor \mu q \rfloor} B_j \leq 1-\bar{\mu}\}$. 
\end{theorem}


We prove Theorem~1 In Section~\ref{sec:scheme}, in which we present a unified coded scheme that jointly designs the storage and the data shuffling, which achieves the latency in (\ref{eq:latency}) and the communication load in (\ref{eq:load}). 


\begin{remark}
The Minimum Latency Code and the Minimum Bandwidth Code correspond to $q = \lceil \frac{1}{\mu}\rceil$ and $q=K$, and achieve the two end points $(\mathbb{E}\{S_{(\lceil \frac{1}{\mu}\rceil)}\}, N-N/\lceil \frac{1}{\mu}\rceil)$ and $(\mathbb{E}\{S_{(K)}\}, N\frac{1-\lfloor \mu K\rfloor/K}{\lfloor \mu K\rfloor})$ respectively.  $\hfill \square$
\end{remark}




\begin{figure}[htbp]
   \centering
   \includegraphics[width=0.3\textwidth]{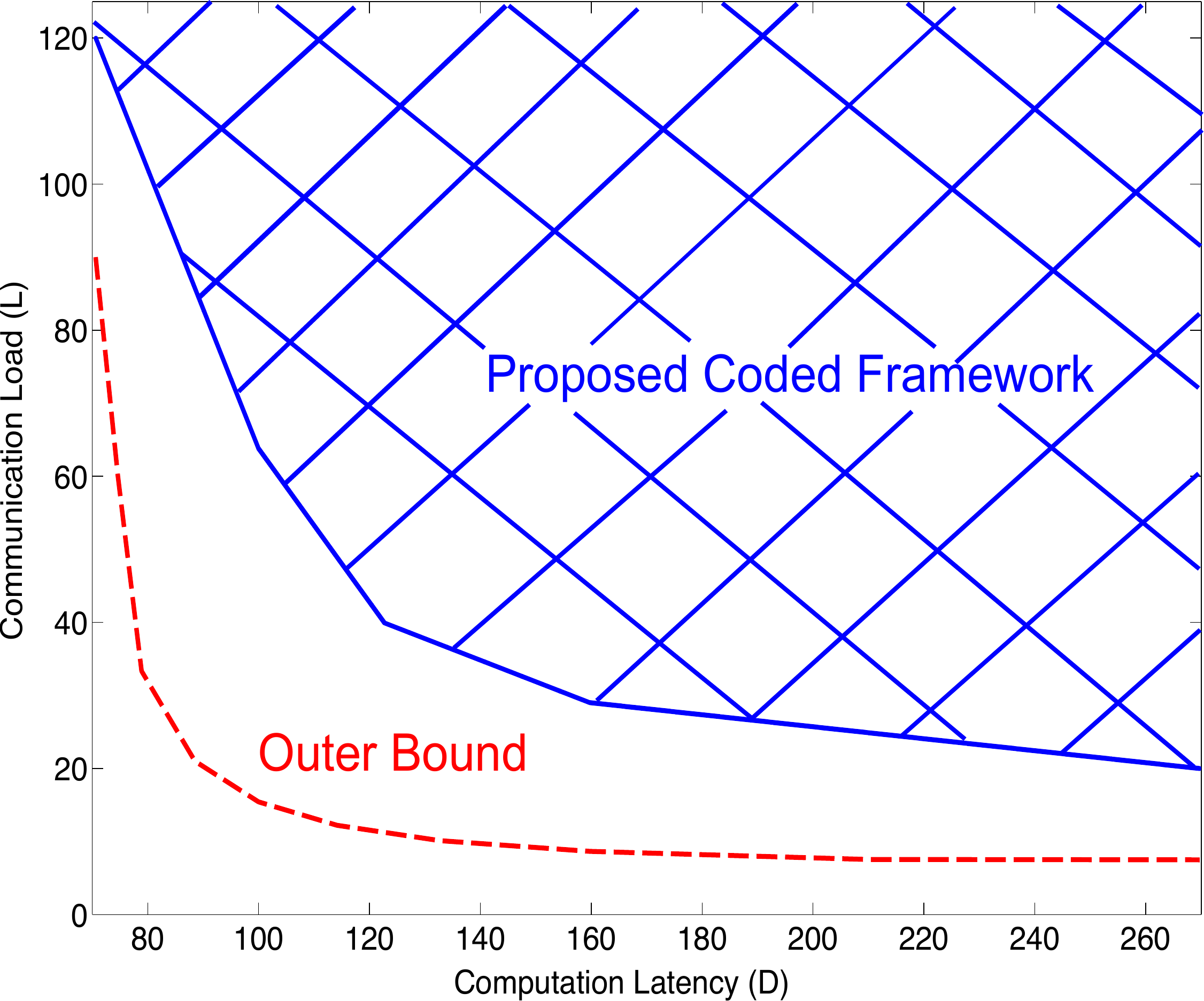}
   \caption{Comparison of the latency-load pairs achieved by the proposed scheme with the outer bound, for computing $N=180$ output vectors using $K=18$ servers each with a storage size $\mu=1/3$, assuming the the distribution function of the Map time in (\ref{eq:dis}).}
   \vspace{-2mm}
   \label{fig:region}
\end{figure}

\begin{remark}
We numerically evaluate in Fig.~\ref{fig:region} the latency-load pairs achieved by the proposed coded framework, for computing $N\!=\!180$ output vectors using $K\!=\!18$ servers each with a storage size $\mu \!=\!1/3$. 
The achieved tradeoff approximately exhibits an inverse-linearly proportional relationship between the latency and the load. For instance, doubling the latency from 120 to 240 results in a drop of the communication load from 43 to 23 by a factor of 1.87.$\hfill \square$
\end{remark}

\begin{remark}
The key idea to achieve $D(q)$ and $L(q)$ in Theorem~1 is to design the concatenation of the MDS code and the repetitive executions of the Map computations, in order to take advantage of both the Minimum Latency Code and the Minimum Bandwidth Code. More specifically, we first generate $\frac{K}{q}m$ MDS-coded rows of ${\bf A}$, and then store each of them $\lfloor \mu q\rfloor$ times across the $K$ servers in a specific pattern. As a result, any subset of $q$ servers would have sufficient amount of intermediate results to reduce the output vectors, and we end the Map phase as soon as the fastest $q$ servers finish their Map computations, achieving the latency in (\ref{eq:latency}).

We also exploit coded multicasting in the Shuffle phase to reduce the communication load. In the load expression (\ref{eq:load}), $B_j$, $j \leq \lfloor \mu q \rfloor$, represents the (normalized) number of coded rows of ${\bf A}$ repeatedly stored/computed at $j$ servers. By multicasting coded packets simultaneously useful for $j$ servers, $B_j$ intermediate values can be delivered to a server with a communication load of $\frac{B_j}{j}$, achieving a coding gain of $j$. We greedily utilize the coding opportunities with a larger coding gain until we get close to satisfying the demand of each server, which accounts for the first term in (\ref{eq:load}). Then the second term results from two follow-up strategies 1) communicate the rest of the demands uncodedly 2) continue coded multicasting with a smaller coding gain (i.e., $j=s_q-1$), which may however deliver more than what is needed for reduction. $\hfill \square$
\end{remark}


\vspace{-2mm}
\begin{theorem}
The latency-load region is contained in the lower convex envelop of the points
\begin{align}
\{(D(q),\bar{L}(q)): q =\lceil \tfrac{1}{\mu}\rceil,\ldots,K\},
\end{align}
in which $D(q)$ is given by (\ref{eq:latency}) and 
\begin{align}
\bar{L}(q) = N\underset{t=1,\ldots,q-1}{\max} \frac{1-\min\{t\mu, 1\}}{\lceil \tfrac{q}{t}\rceil (q-t)}q.\label{eq:lower}
\end{align}
\end{theorem}

We prove Theorem~2 in Section~V, by deriving an information-theoretic lower bound on the minimum required communication load for a given computation latency, using any storage design and data shuffling scheme. 

\vspace{-1.3mm}
\begin{remark}
We numerically compare the outer bound in Theorem~2 and the achieved inner bound in Theorem~1 in Fig.~\ref{fig:region}, from which we make the following observations.
\vspace{-1.2mm}
\begin{itemize}[leftmargin=4mm]
\item At the minimum latency point, i.e., $q=1 /\mu=3$ servers finish the Map computations, the proposed coded scheme achieves $1.33 \times$ of the minimum communication load. In general, when $q= 1/\mu \in \mathbb{N}$, the lower bound in Theorem~2 $\bar{L}(\frac{1}{\mu}) = N/ \lceil \frac{q}{t}\rceil |_{t=q-1} = N/\lceil \frac{1}{1-\mu}\rceil = \frac{N}{2}$. The proposed coded scheme, or Minimum Latency Code in this case, achieves the load $L(\frac{1}{\mu}) =N(1-\mu)$. Thus the proposed scheme always achieves the lower bound to within a factor of 2 at the minimum latency point. 

\item At the point with the maximum latency, i.e., all $K=18$ servers finish the Map computations, the proposed coded scheme achieves $2.67 \times$ of the lower bound on the minimum communication load. In general for $q=K$ and $\mu K \in \mathbb{N}$, we demonstrate in Appendix that the proposed coded scheme, or Minimum Bandwidth Code in this case, achieves a communication load $L(K) = N(1-\mu)/(\mu K)$ to within a factor of $3+\sqrt{5}$ of the lower bound $\bar{L}(K)$. 

\item For the intermediate latency from 70 to 270, the communication load achieved by the proposed scheme is within a multiplicative gap of at most $4.2 \times$ from the lower bound. In general, a complete characterization of the latency-load region (or an approximation to within a constant gap for all system parameters) remains open.$\hfill \square$
\end{itemize}
\end{remark}


\section{Proposed Coded Framework}\label{sec:scheme}
In this section, we prove Theorem~1 by proposing and analyzing a general coded framework that achieves the latency-load pairs in (\ref{eq:pair}). We first demonstrate the key ideas of the proposed scheme through the following example, and then give the general description of the scheme.

\subsection{Example: $m=20$, $N=12$, $K=6$ and $\mu =\frac{1}{2}$.}
We have a problem of multiplying a matrix ${\bf A} \in \mathbb{F}_{2^T}^{m \times n}$ of $m=20$ rows with $N=12$ input vectors ${\bf x}_1,\ldots,{\bf x}_{12}$ to compute $12$ output vectors ${\bf y}_1={\bf A}{\bf x}_1\ldots,{\bf y}_{12}={\bf A}{\bf x}_{12}$, using $K=6$ servers each with a storage size $\mu =\frac{1}{2}$. 


We assume that we can afford to wait for $q=4$ servers to finish their computations in the Map phase, and we describe the proposed storage design and shuffling scheme.

\begin{figure}[htbp]
   \centering
   \includegraphics[width=0.48\textwidth]{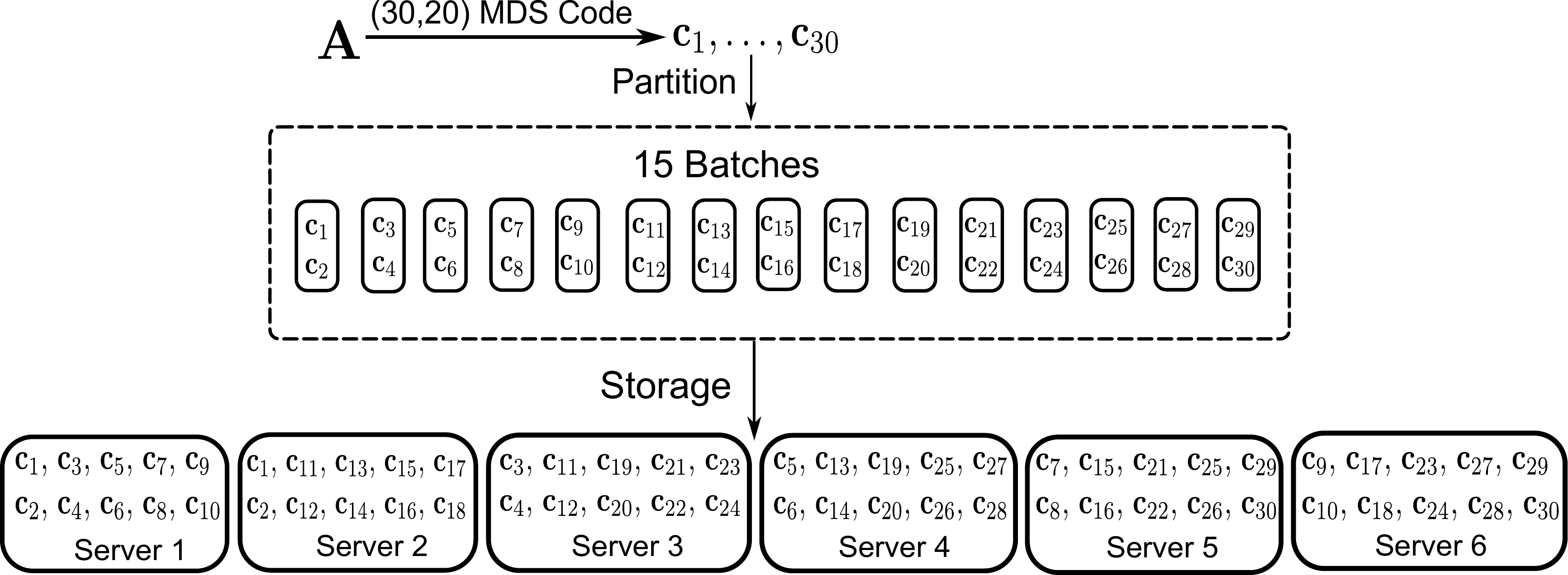}
   \caption{Storage Design when the Map phase is terminated when $4$ servers have finished the computations.}
   \label{fig:example-storage}
\end{figure}

\noindent {\bf Storage Design.} As illustrated in Fig~\ref{fig:example-storage}, we first independently generate $30$ random linear combinations ${\bf c}_1,\ldots,{\bf c}_{30} \in \mathbb{F}_{2^T}^n$ of the $20$ rows of ${\bf A}$, achieving a $(30,20)$ MDS code of the rows of ${\bf A}$. Then we partition these coded rows ${\bf c}_1,\ldots,{\bf c}_{30}$ into $15$ batches each of size $2$, and store every batch of coded rows at a unique pair of servers. 

WLOG, due to the symmetry of the storage design, we assume that Servers $1$, $2$, $3$ and $4$ are the first $4$ servers that finish their Map computations. Then we assign the Reduce tasks such that Server $k$ reduces the output vectors ${\bf y}_{3(k-1)+1}$, ${\bf y}_{3(k-1)+2}$ and ${\bf y}_{3(k-1)+3}$, for all $k \in \{1,\ldots,4\}$. 

After the Map phase, Server~1 has computed the intermediate values $\{{\bf c}_1{\bf x}_j, \ldots,{\bf c}_{10}{\bf x}_j: j=1,\ldots,12\}$. For Server~1 to recover ${\bf y}_1 = {\bf A}{\bf x}_1$, it needs any subset of 10 intermediate values ${\bf c}_i{\bf x}_1$ with $i \in \{11,\ldots,30\}$ from Server $2$, $3$ and $4$ in the Shuffle phase. Similar data demands hold for all 4 servers and the output vectors they are reducing. Therefore, the goal of the Shuffle phase is to exchange these needed intermediate values to accomplish successful reductions.

\begin{figure}[htbp]
   \centering
   \includegraphics[width=0.4\textwidth]{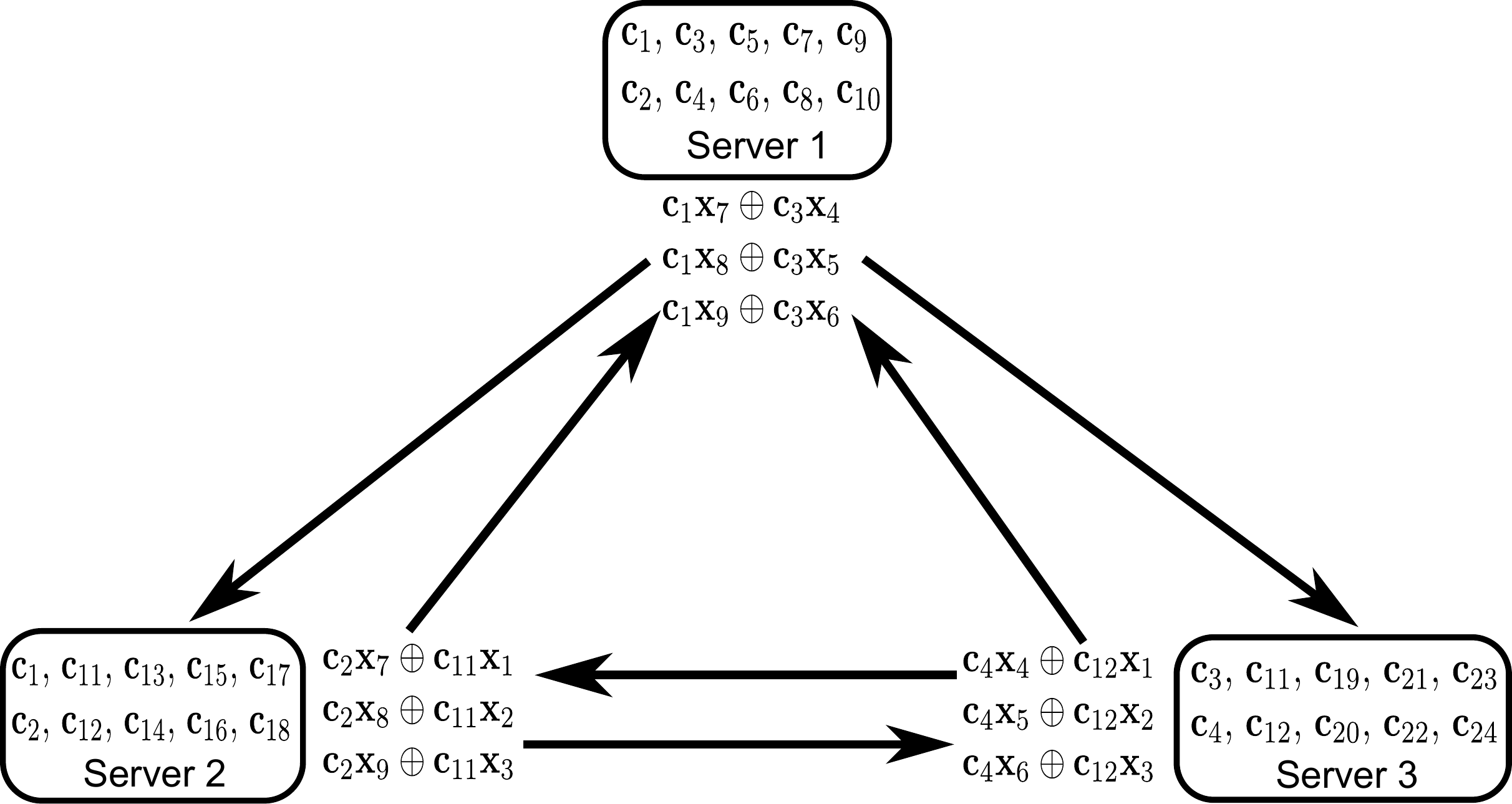}
   \caption{Multicasting 9 coded intermediate values across Servers~1, 2 and 3. Similar coded multicast communications are performed for another 3 subsets of 3 servers.} 
   \label{fig:example-shuffle}
\end{figure}

\noindent {\bf Coded Shuffle.} We first group the 4 servers into 4 subsets of size 3 and perform coded shuffling within each subset. 
We illustrate the coded shuffling scheme for Servers $1$, $2$ and $3$ in Fig.~\ref{fig:example-shuffle}. Each server multicasts $3$ bit-wise XORs, denoted by $\oplus$, of the locally computed intermediate values to the other two. The intermediate values used to create the multicast messages are the ones known exclusively at two servers and needed by another one. After receiving $2$ multicast messages, each server recovers $6$ needed intermediate values. For instance, Server~1 recovers ${\bf c}_{11}{\bf x}_1$, ${\bf c}_{11}{\bf x}_2$ and ${\bf c}_{11}{\bf x}_3$ by canceling ${\bf c}_{2}{\bf x}_7$, ${\bf c}_{2}{\bf x}_8$ and ${\bf c}_{2}{\bf x}_9$ respectively, and then recovers ${\bf c}_{12}{\bf x}_1$, ${\bf c}_{12}{\bf x}_2$ and ${\bf c}_{12}{\bf x}_3$ by canceling ${\bf c}_{4}{\bf x}_4$, ${\bf c}_{4}{\bf x}_5$ and ${\bf c}_{4}{\bf x}_6$ respectively.

Similarly, we perform the above coded shuffling in Fig.~\ref{fig:example-shuffle} for another $3$ subsets of $3$ servers. After coded multicasting within the $4$ subsets of $3$ servers, each server recovers $18$ needed intermediate values (6 for each of the output vector it is reducing). As mentioned before, since each server needs a total of $3\times (20-10)=30$ intermediate values to reduce the 3 assigned output vectors, it needs another $30-18=12$ after decoding all multicast messages. We satisfy the residual data demands by simply having the servers unicast enough (i.e., $12 \times 4=48$) intermediate values for reduction. Overall, $9\times 4+48= 84$ (possibly coded) intermediate values are communicated, achieving a communication load of $L= 4.2$. 

\subsection{General Scheme}
We first describe the storage design, Map phase computation and the data shuffling scheme that achieves the latency-load pairs $(D(q),L(q))$ in (\ref{eq:pair}), for all $q \in \{\lceil \frac{1}{\mu} \rceil, \ldots,K\}$. Given these achieved pairs, we can ``memory share'' across them to achieve their lower convex envelop as stated in Theorem~1.

For ease of exposition, we assume that $\mu q \in \mathbb{N}$. Otherwise we can replace $\mu$ with $\bar{\mu}=\frac{\lfloor \mu q \rfloor}{q}$, and apply the proposed scheme for a storage size of $\bar{\mu}$.
  
\begin{figure}[htbp]
   \centering
   \includegraphics[width=0.35\textwidth]{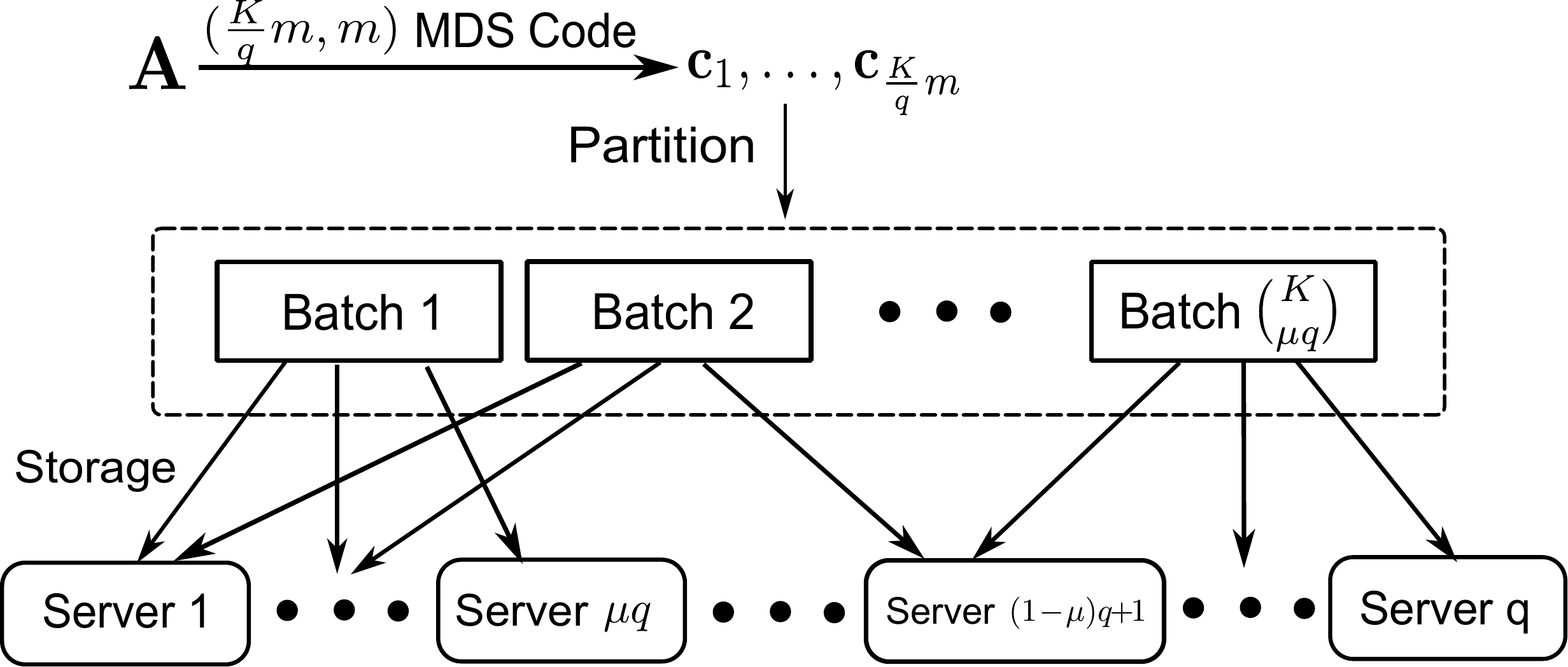}
   \caption{General MDS coding and storage design.}
   \label{fig:storage}
\end{figure}

\noindent {\bf Storage Design.} 
We first use a $(\frac{K}{q}m,m)$ MDS code to encode the $m$ rows of matrix ${\bf A}$ into $\frac{K}{q}m$ coded rows ${\bf c}_1\ldots,{\bf c}_{\frac{K}{q}m}$ (e.g., $\frac{K}{q}m$ random linear combinations of the rows of ${\bf A}$). Then as shown in Fig.~\ref{fig:storage}, we evenly partitioned the $\frac{K}{q}m$ coded rows into ${K \choose \mu q}$ disjoint batches, each containing a subset of $\frac{m}{\frac{q}{K} {K \choose \mu q}}$ coded rows. \footnote{We focus on matrix multiplication problems for large matrices, and assume that $m \gg \frac{q}{K} {K \choose \mu q}$, for all $q \in \{\frac{1}{\mu},\ldots,K\}$.} Each batch, denoted by ${\cal B}_{\cal T}$, is labelled by a unique subset $\mathcal{T} \subset \{1,\ldots,K\}$ of size $|{\cal T}|=\mu q$. That is
\begin{align}
\{1,\ldots,\tfrac{K}{q}m\} = \{\mathcal{B}_{\cal T}: {\cal T} \subset \{1,\ldots,K\}, |{\cal T}|=\mu q \}.
\end{align}

Server~$k$, $k \in \{1,\ldots,K\}$ stores the coded rows in $\mathcal{B}_{\cal T}$ as the rows of ${\bf U}_k$ if $k \in \mathcal{T}$. 

In the above example, $q=4$, and $\frac{K}{q}m= \frac{6}{4} \times 20=30$ coded rows of ${\bf A}$ are partitioned into ${K \choose \mu q}={6 \choose 2}=15$ batches each containing $\frac{30}{15}=2$ coded rows. Every node is in $5$ subsets of size two, thus storing $5 \times 2=10$ coded rows of ${\bf A}$.

\noindent {\bf Map Phase Execution.}  Each server computes the inner products between each of the locally stored coded rows of ${\bf A}$ and each of the input vectors, i.e., Server $k$ computes ${\bf c}_i{\bf x}_j$ for all $j=1,\ldots,N$, and all $i \in \{{\cal B}_{\cal T}: k \in {\cal T}\}$. We wait for the fastest $q$ servers to finish their Map computations before halting the Map phase, achieving a computation latency $D(q)$ in (\ref{eq:latency}). We denote the set of indices of these servers as ${\cal Q}$.

The computation then moves on exclusively over the $q$ servers in ${\cal Q}$, each of which is assigned to reduce $\frac{N}{q}$ out of the $N$ output vectors ${\bf y}_1={\bf A}{\bf x}_1,\ldots,{\bf y}_N={\bf A}{\bf x}_N$. 

For a feasible shuffling scheme to exist such that the Reduce phase can be successfully carried out, every subset of $q$ servers (since we cannot predict which $q$ servers will finish first) should have collectively stored at least $m$ distinct coded rows ${\bf c}_i$ for $i \in \{1,\ldots,\frac{K}{q}m\}$. Next, we explain how our proposed storage design meets this requirement. First, the $q$ servers in ${\cal Q}$ collectively provide a storage size equivalent to $\mu q m$ rows. 
Then since each coded row is stored by $\mu q$ out of all $K$ servers, it can be stored by at most $\mu q$ servers in ${\cal Q}$, and thus servers in ${\cal Q}$ collectively store at least $\frac{\mu q m}{\mu q}=m$ distinct coded rows.

\noindent {\bf Coded Shuffle.} For ${\cal S}\subset {\cal Q}$ and $k \in {\cal Q} \backslash {\cal S}$, we denote the set of intermediate values needed by Server $k$ and known \emph{exclusively} by the servers in $\mathcal{S}$ as $\mathcal{V}_{\mathcal{S}}^{k}$. More formally:
\begin{equation}\label{eq:V}
\mathcal{V}_{\mathcal{S}}^{k} \triangleq \{{\bf c}_i{\bf x}_j: j \in {\cal W}_k, i \in \{{\cal B}_{\cal T}: {\cal T} \cap {\cal Q}={\cal S}\}\}.
\end{equation}

Due to the proposed storage design, for a particular ${\cal S}$ of size $j$, $\mathcal{V}_{\mathcal{S}}^{k}$ contains $\frac{N}{q}\cdot\frac{{K-q \choose \mu q-j}m}{\frac{q}{K} {K \choose \mu q}}$ intermediate values.

In the above example, we have $\mathcal{V}_{\{2,3\}}^1=\{{\bf c}_{11}{\bf x}_j,{\bf c}_{12}{\bf x}_j: j=1,2,3\}$, $\mathcal{V}_{\{1,3\}}^2=\{{\bf c}_{3}{\bf x}_j,{\bf c}_{4}{\bf x}_j: j=4,5,6\}$, and $\mathcal{V}_{\{1,2\}}^3=\{{\bf c}_{1}{\bf x}_j,{\bf c}_{2}{\bf x}_j: j=7,8,9\}$. 

In the Shuffle phase, servers in ${\cal Q}$ create and multicast coded packets that are simultaneously useful for multiple other servers, until every server in ${\cal Q}$ recovers at least $m$ intermediate values for each of the output vectors it is reducing. The proposed shuffling scheme is \emph{greedy} in the sense that every server in ${\cal Q}$ will always try to multicast coded packets simultaneously useful for the largest number of servers.

The proposed shuffle scheme proceeds as follows. For each $j\!=\!\mu q, \mu q-1,\ldots,s_q$, where $s_q \!\triangleq\! \inf \{s: \! \sum_{j=s}^{\mu q} \! \frac{{q-1 \choose j}{K-q \choose \mu q-j}}{\frac{q}{K} {K \choose \mu q}} \!\leq\! 1\!-\!\mu\}$, and every subset $\mathcal{S} \!\subseteq\! {\cal Q}$ of size $j\!+\!1$:
\begin{enumerate}[leftmargin=5mm]
\item For each $k \in \mathcal{S}$, we evenly and arbitrarily split $\mathcal{V}_{\mathcal{S}\backslash \{k\}}^{k}$ into $j$ disjoint segments $\mathcal{V}^{k}_{\mathcal{S}\backslash \{k\}} \!=\! \{ \mathcal{V}_{\mathcal{S} \backslash \{k\},i}^{k}\!:\! i \in {\cal S} \backslash \{k\}\}$, and associate the segment $\mathcal{V}_{\mathcal{S} \backslash \{k\},i}^{k}$ with the server $i \in {\cal S} \backslash \{k\}$. 
\item Server $i$, $i \in \mathcal{S}$, multicasts the bit-wise XOR, denoted by $\oplus$, of all the segments associated with it in ${\cal S}$, i.e., Server $i$ multicasts $ \underset{k \in \mathcal{S} \backslash \{i\}} \oplus \mathcal{V}^{k}_{\mathcal{S}\backslash \{k\},i}$ to the other servers in ${\cal S} \backslash \{i\}$.
\end{enumerate}

For every pair of servers $k$ and $i$ in ${\cal S}$, 
since Server $k$ has computed locally the segments $\mathcal{V}^{k'}_{\mathcal{S}\backslash \{k'\},i}$ for all $k' \in \mathcal{S} \backslash \{i,k\}$, it can cancel them from the message $\underset{k \in \mathcal{S} \backslash \{i\}}\oplus \mathcal{V}^{k}_{\mathcal{S}\backslash \{k\},i}$ sent by Server $i$, and recover the intended segment $\mathcal{V}^{k}_{\mathcal{S}\backslash \{k\},i}$.

For each $j$ in the above coded shuffling scheme, each server in ${\cal Q}$ recovers ${q-1 \choose j}\frac{{K-q \choose \mu q-j}m}{\frac{q}{K} {K \choose \mu q}}$
intermediate values for each of the output vectors it is reducing. Therefore, $j=s_q+1$ is the smallest size of the subsets in which the above coded multicasting needs to be performed, before enough number of intermediate values for reduction are delivered.

In each subset ${\cal S}$ of size $j$, since each server $i \in {\cal S}$ multicasts a coded segment of size $\frac{|{\cal V}^k_{{\cal S} \backslash \{k\}}|}{j}$ for some $k \neq i$, 
the total communication load so far, for $B_j = \frac{{q-1 \choose j}{K-q \choose \mu q-j}}{\frac{q}{K} {K \choose \mu q}}$, is
\begin{align}
\sum_{j=s_q}^{\mu q}{q \choose j+1}\frac{j+1}{j}\cdot \frac{N}{q} \cdot \frac{{K-q \choose \mu q-j}}{\frac{q}{K} {K \choose \mu q}}=\sum_{j=s_q}^{\mu q} N \frac{B_j}{j},
\end{align}

Next, we can continue to finish the data shuffling in two different ways. The first approach is to have the servers in ${\cal Q}$ communicate with each other uncoded intermediate values, until every server has exactly $m$ intermediate values for each of the output vector it is responsible for. Using this approach, we will have a total communication load of 
\begin{align}
L_1=\sum_{j=s_q}^{\mu q} N \tfrac{B_j}{j} + N(1-\mu-\sum_{j=s_q}^{\mu q}B_j).
\end{align}

The second approach is to continue the above 2 steps for $j=s_q-1$. Using this approach, we will have a total communication load of 
$L_2=\sum_{j=s_q-1}^{\mu q} N \frac{B_j}{j}$.

Then we take the approach with less communication load, and achieve $L(q)=\min\{L_1,L_2\}$. 

\begin{remark}
The ideas of efficiently creating and exploiting coded multicasting opportunities have been introduced in caching problems~\cite{maddah2014fundamental,maddah2013decentralized,ji2014fundamental}. In this section, we illustrated how to create and utilize  such coding opportunities in distributed computing to slash the communication load, when facing with straggling servers. 
$\hfill \square$
\end{remark}

\section{Converse}\label{sec:converse}
In this section, we prove the outer bound on the latency-load region in Theorem~2. 

We start by considering a distributed matrix multiplication scheme that stops the Map phase when $q$ servers have finished their computations. For such scheme, as given by (\ref{eq:latency}), the computation latency $D(q)$ is the expected value of the $q$th order statistic of the Map computation times at the $K$ servers. WLOG, we can assume that Servers $1,\ldots,q$ first finish their Map computations, and they will be responsible for reducing the $N$ output vectors ${\bf y}_1,\ldots,{\bf y}_N$. 

To proceed, we first partition the ${\bf y}_1,\ldots,{\bf y}_N$ into $q$ groups ${\cal G}_1,\ldots,{\cal G}_q$ each of size $N/q$, and define the \emph{output assignment}
\begin{align}
{\cal A} = \left({\cal W}_1^{\cal A},{\cal W}_2^{\cal A}\ldots,{\cal W}_q^{\cal A}\right),
\end{align} 
where ${\cal W}_k^{\cal A}$ denotes the group of output vectors reduced by Server $k$ in the output assignment ${\cal A}$.

Next we choose an integer $t \in \{1,\ldots,q-1\}$, and consider the following $\lceil \frac{q}{t} \rceil$ output assignments which are circular shifts of $\left({\cal G}_1,\ldots,{\cal G}_q\right)$ with step size $t$,
\begin{equation}\label{eq:assign}
\begin{aligned}
\mathcal{A}_1 &= \left({\cal G}_1,{\cal G}_2,\ldots,{\cal G}_q\right),\\
\mathcal{A}_2 &= \left({\cal G}_{t+1},\ldots,{\cal G}_q, {\cal G}_1,\ldots, {\cal G}_t\right),\\
& \vdots\\
\mathcal{A}_{\lceil \frac{q}{t} \rceil} &= \left({\cal G}_{(\lceil\frac{q}{t} \rceil \!-\!1)t+1},\ldots,{\cal G}_q, {\cal G}_1,\ldots,{\cal G}_{(\lceil \frac{q}{t} \rceil-1) t}\right).
\end{aligned}
\end{equation}

\begin{remark}\label{independence}
We note that by the Map computation in (\ref{eq:map}), at each server all the input vectors ${\bf x}_1,\ldots,{\bf x}_N$ are multiplied by the same matrix (i.e., ${\bf U}_k$ at Server~$k$). Therefore, for the same set of $q$ servers and their storage contents, a feasible data shuffling scheme for one of the above output assignments is also feasible for all other $\lceil \frac{q}{t} \rceil-1$ assignments by relabelling the output vectors. As a result, the minimum communication loads for all of the above output assignments are identical. $\hfill \square$
\end{remark}

For a shuffling scheme admitting an output assignment ${\cal A}$, we denote the message sent by Server $k \in \{1,\ldots,q\}$ as $X_k^{\mathcal{A}}$, with a size of $R_{k}^{\mathcal{A}}mT$ bits.

\begin{figure}[htbp]
   \centering
   \includegraphics[width=0.48\textwidth]{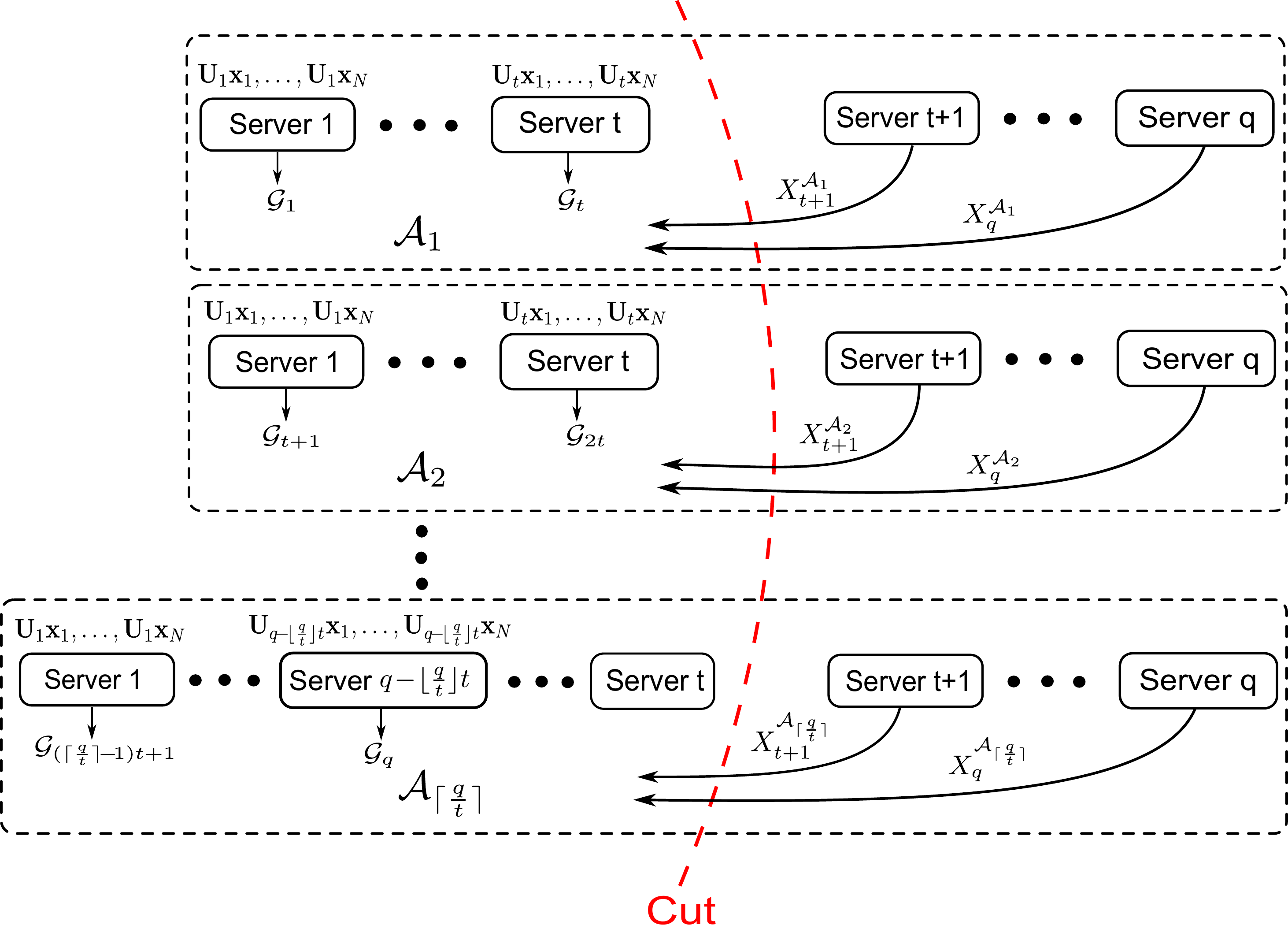}
   \caption{Cut-set of Servers $1,\ldots,t$ for the compound setting consisting of the $\lceil \frac{q}{t} \rceil$ output assignments in (\ref{eq:assign}).}
   \label{fig:compound}
\end{figure}

Now we focus on the Servers $1,\ldots,t$ and consider the compound setting that includes all $\lceil \frac{q}{t} \rceil$ output assignments in (\ref{eq:assign}). We observe that as shown in Fig.~\ref{fig:compound}, in this compound setting, the first $t$ servers should be able to recover all output vectors $({\bf y}_1\ldots,{\bf y}_N) = ({\cal G}_1,\ldots,{\cal G}_q)$ using their local computation results $\{{\bf U}_k{\bf x}_1,\ldots,{\bf U}_k{\bf x}_N:k=1,\ldots,t\}$ and the received messages in all the output assignments $\{X_k^{{\cal A}_1},\ldots,X_k^{{\cal A}_{\lceil \frac{q}{t}\rceil}}:k=t+1,\ldots,q\}$. Thus we have the following cut-set bound for the first $t$ servers.
\begin{equation}
rank \left( \begin{bmatrix} {\bf U}_1 \\ {\bf U}_2  \\ \vdots \\ {\bf U}_{t} \end{bmatrix} \right) NT + \sum \limits_{j=1}^{\lceil \frac{q}{t}\rceil} \sum \limits_{k=t+1}^{K} R_{k}^{\mathcal{A}_j}mT \geq NmT.
\end{equation}

Next we consider $q$ subsets of servers each with size $t$: $\mathcal{N}_i \triangleq \{i, (i+1), \ldots, (i+t-1)\}$, $i = 1,\ldots,q$, where the addition is modular $q$. Similarly, we have the following cut-set bound for ${\cal N}_i$:
\begin{equation}
rank \left( \begin{bmatrix} {\bf U}_i \\ {\bf U}_{i+1}  \\ \vdots \\ {\bf U}_{i+t-1} \end{bmatrix} \right) NT + \sum \limits_{j=1}^{\lceil \frac{q}{t}\rceil} \sum \limits_{k \notin \mathcal{N}_i} R_{k}^{\mathcal{A}_j}mT \geq NmT.
\end{equation}

Summing up these $q$ cut-set bounds, we have
\begin{align}
NT\! \sum \limits_{i=1}^q rank \!\! \left(\! \begin{bmatrix} {\bf U}_i \\ {\bf U}_{i+1}  \\ \vdots \\ {\bf U}_{i+t-1} \end{bmatrix} \!\right)& \!\! + \!  \sum \limits_{i=1}^q \sum \limits_{j=1}^{\lceil \frac{q}{t}\rceil}  \sum \limits_{k \notin \mathcal{N}_i} \!\!R_{k}^{\mathcal{A}_j}mT  \geq qNmT, \\
\Rightarrow \sum \limits_{j=1}^{\lceil \frac{q}{t}\rceil}  \sum \limits_{i=1}^q \sum \limits_{k \notin \mathcal{N}_i} R_{k}^{\mathcal{A}_j} \geq& qN-qN\min\{\mu t,1\}.\\
\Rightarrow   \lceil \tfrac{q}{t}\rceil (q-t)L  \overset{(a)}{\geq}& (1-\min\{t\mu, 1\})qN, \label{eq:sumup}
\end{align}
where (a) results from the fact mentioned in Remark~\ref{independence} that the communication load is independent of the output assignment.

Since (\ref{eq:sumup}) holds for all $t=1,\ldots,q-1$, we have
\begin{align}
L  \geq \bar{L}(q) =N\underset{t=1,\ldots,q-1}{\max} \frac{1-\min\{t\mu, 1\}}{\lceil \tfrac{q}{t}\rceil (q-t)}q.
\end{align}

We assume that the Map phase terminates when $q$ servers finish the computations with probability $P(q)$, for all $q \in \{\lceil \frac{1}{\mu}\rceil, \ldots,K\}$, then the communication load for a latency $\mathbb{E}_{q}(D(q))$ that is a convex combination of $\{\mathbb{E}\{S_{(q)}\}: q=\lceil \frac{1}{\mu}\rceil, \ldots,K\}$, is lower bounded by $\mathbb{E}_{q}(\bar{L}(q))$ that is the same convex combination of $\{\bar{L}(q): q=\lceil \frac{1}{\mu}\rceil, \ldots,K)\}$. Considering all distributions of $q$, we achieve all points on the lower convex envelop of the points $\{(\mathbb{E}\{S_{(q)}\}, \bar{L}(q)): q=\lceil \frac{1}{\mu}\rceil, \ldots,K\}$, as an outer bound on the latency-load region.

\appendix
In this appendix, we prove that when all $K$ servers finish their Map computations, i.e., ${\cal Q}=\{1,\ldots,K\}$ and we operate at the point with the maximum latency, the communication load achieved by the proposed coded scheme (or the Minimum Bandwidth Code) is within a constant multiplicative factor of the lower bound on the communication load in Theorem~2. More specifically,
\begin{align}
\frac{L(K)}{\bar{L}(K)} < 3+\sqrt{5},
\end{align} 
when $\mu K$ is an integer,\footnote{This always holds true for large $K$.} where $L(K)$ and $\bar{L}(K)$ are respectively given by (\ref{eq:load}) and (\ref{eq:lower}).

\noindent \emph{Proof.} For $\mu K \in \mathbb{N}$, we have $L(K)=N\frac{1-\mu}{\mu K}$, and 
\begin{align}
\frac{L(K)}{\bar{L}(K)}=\frac{\frac{1-\mu}{\mu K}}{\underset{t=1,\ldots,K-1}{\max} \tfrac{1-\min\{t\mu, 1\}}{\lceil \frac{K}{t}\rceil (K-t)}K}. \label{eq:constGap1}
\end{align}

We proceed to bound the RHS of (\ref{eq:constGap1}) in the following two cases:

\noindent 1) $1 \leq \frac{1}{\mu} \leq 3+\sqrt{5}$.

We set $t = 1$ in (\ref{eq:constGap1}) to have 
\begin{align}
\frac{L(K)}{\bar{L}(K)} \leq \frac{\tfrac{1-\mu}{\mu K}}{\tfrac{1-\mu}{K-1}} < \frac{1}{\mu} \leq 3+\sqrt{5}. \label{eq:constGap2}
\end{align}

\noindent 2) $\frac{1}{\mu} > 3+\sqrt{5}$.

Since $\mu K \geq 1$, we have $K-1 \geq \lceil \frac{K}{2}\rceil \geq \lceil \frac{1}{2 \mu}\rceil$. 

In this case, we set $t = \lceil \frac{1}{2\mu}\rceil$ in (\ref{eq:constGap1}) to have
\begin{align}
\frac{L(K)}{\bar{L}(K)} &\leq  \frac{(1-\mu) \lceil \frac{K}{\lceil \frac{1}{2 \mu}\rceil}\rceil (K-\lceil \frac{1}{2 \mu}\rceil)}{\mu K^2(1- \mu \lceil \frac{1}{2\mu}\rceil)}\\
& \leq  \frac{2(1-\mu) (K-\lceil \frac{1}{2 \mu}\rceil)}{K(1- \mu \lceil \frac{1}{2\mu}\rceil)} <   \frac{2(1-\mu)}{1- \mu \lceil \frac{1}{2\mu}\rceil} \\
& \leq   \frac{2(1-\mu)}{1- \mu (\frac{1}{2\mu}+1)}\\
& = 4 + \frac{4}{\frac{1}{\mu}-2}< 3+ \sqrt{5}, \label{eq:constGap3}
\end{align}

Comparing (\ref{eq:constGap2}) and (\ref{eq:constGap3}) completes the proof. $\hfill \blacksquare$

\bibliographystyle{IEEEtran}
\bibliography{ref-abb}

\end{document}